\documentclass[12pt]{iopart}
\usepackage{graphicx}
\usepackage{color}

\begin{document}
\title{On the superconductivity of TiNCl and ZrNCl: A local bonding perspective}
\author{Lukas Muechler}
\ead{muechler@princeton.edu}
\address{Department of Chemistry, Princeton University, Princeton New Jersey 08544, USA.}
\address{Max-Planck-Institut f\"ur Chemische Physik fester Stoffe, 01187 Dresden, Germany.}
\author{Leslie M. Schoop}
\address{Department of Chemistry, Princeton University, Princeton New Jersey 08544, USA.}
\address{Max-Planck-Institut f\"ur Chemische Physik fester Stoffe, 01187 Dresden, Germany.}
\author{Claudia Felser}
\address{Max-Planck-Institut f\"ur Chemische Physik fester Stoffe, 01187 Dresden, Germany.}

\begin{abstract}
We analyze the superconductors TiNCl and ZrNCl from a local bonding perspective. Although TiNCl crystallizes in an orthorhombic structure and ZrNCl crystallizes in a hexagonal structure, both compounds show significant structural similarities, for example that both consist of layered metal-Nitrogen networks.
The local bonding in those two structures is very similar, giving rise to a dispersive conduction band mostly consisting of metal-\textit{d}-states. Upon doping both compounds show structural changes, which lead to short metal-metal distances, indicating a bonding
interaction that might be important for the appearance of superconductivity in these systems.
We furthermore draw analogies to other superconductors that are close to a charge density wave instability around a \textit{d}$^1$-configuration and offer a 
different perspective on this class of superconductors, which show non-BCS-like superconductivity.
\end{abstract}

% Uncomment for PACS numbers
%
% Uncomment for keywords
%\vspace{2pc}
%\noindent{\it Keywords}: XXXXXX, YYYYYYYY, ZZZZZZZZZ
%
% Uncomment for Submitted to journal title message
% \submitto{\JPCM}
%
% Uncomment if a separate title page is required
%\maketitle
% 
% For two-column output uncomment the next line and choose [10pt] rather than [12pt] in the \documentclass declaration
%\ioptwocol
%

\section{Introduction}
The discovery of superconductivity up to 25.5 K in electron doped HfNCl has
attracted a lot of attention, due to its
2-dimensional electronic structure and its high critical temperature T$_c$ comparable to the "other"
high T$_c$ materials such as BaBiO$_3$, La$_2$CuO$_4$ or MgB$_2$.
\cite{yamanaka1998superconductivity}
Shortly after the discovery, the isolectric analogues of Zr have been
synthesized and found to be superconducting upon electron doping as well with
T$_c$s of the order of 15 K.\cite{yamanaka2010,schurz2011}
Only recently the Ti compounds, crystallizing in a different layered structure,
have also been found to superconduct with a T$_c$
of about 15 K. \cite{yamanaka2009}
The mechanism of superconductivity in this class of compounds remains
unexplained. Spin-fluctuation mediated pairing has been suggested as a possible mechanism,
due to the absence of a strong isotope effect, a low density of states at the Fermi level and an
unusual behavior of T$_c$ with doping, with the consensus that the
superconducting state is not BCS-like. In contrast to the dome-like structure of T$_c$ vs. doping level in the other high-T$_c$ materials, T$_c$ in the compounds discussed here is nearly independent of
the doping level after a the appearance of superconductivity at a critical doping.
\cite{PhysRevB.67.100509,kitora2007probing,PhysRevLett.97.107001,
PhysRevB.86.054513}
Only recently however there has been the proposal that electronic correlations play an
important role and that - within a reasonable range of
parameters calculated by \textit{ab-inito} methods - 
T$_c$ can be calculated quite accurately based on a phononic theory.
\cite{Kotliar2013} \\
Most of the studies performed on this class of compounds have focused on the
superconducting phase, especially on the nature of the superconducting pairing
interaction. Little theoretical attention has been paid on the structure to property relationship in these materials, with a few exceptions.
\cite{istomin1999,woodward1998electronic} 
Due to the air-sensitivity of the samples and the rather difficult synthesis,
the Ti-based compounds have not been explored as much as the Zr,Hf -based
compounds. It has been
suggested based on band structure and model calculations that the Ti-based
compounds are different from their Zr and Hf analogues.
\cite{PhysRevB.83.014509} \\
In this paper we are going to show that this is not necessarily the case, based on density functional theory (DFT) calculations. We will show that both ZrNCl and TiNCl have similar features in their electronic structure and 
show their relation to other superconductors close to a \textit{d}$^1$ charge density wave instability.

\section{Crystal structures}
%%%%%%%%%%%%%%%%%%%%%%%%%%%%%%%%%%%%%%%%%%%%%%%%%%%%%%%%%%%%%%%%%%%%%%
\begin{figure}[!htbp]
\centering
  \includegraphics[width=0.95\textwidth]{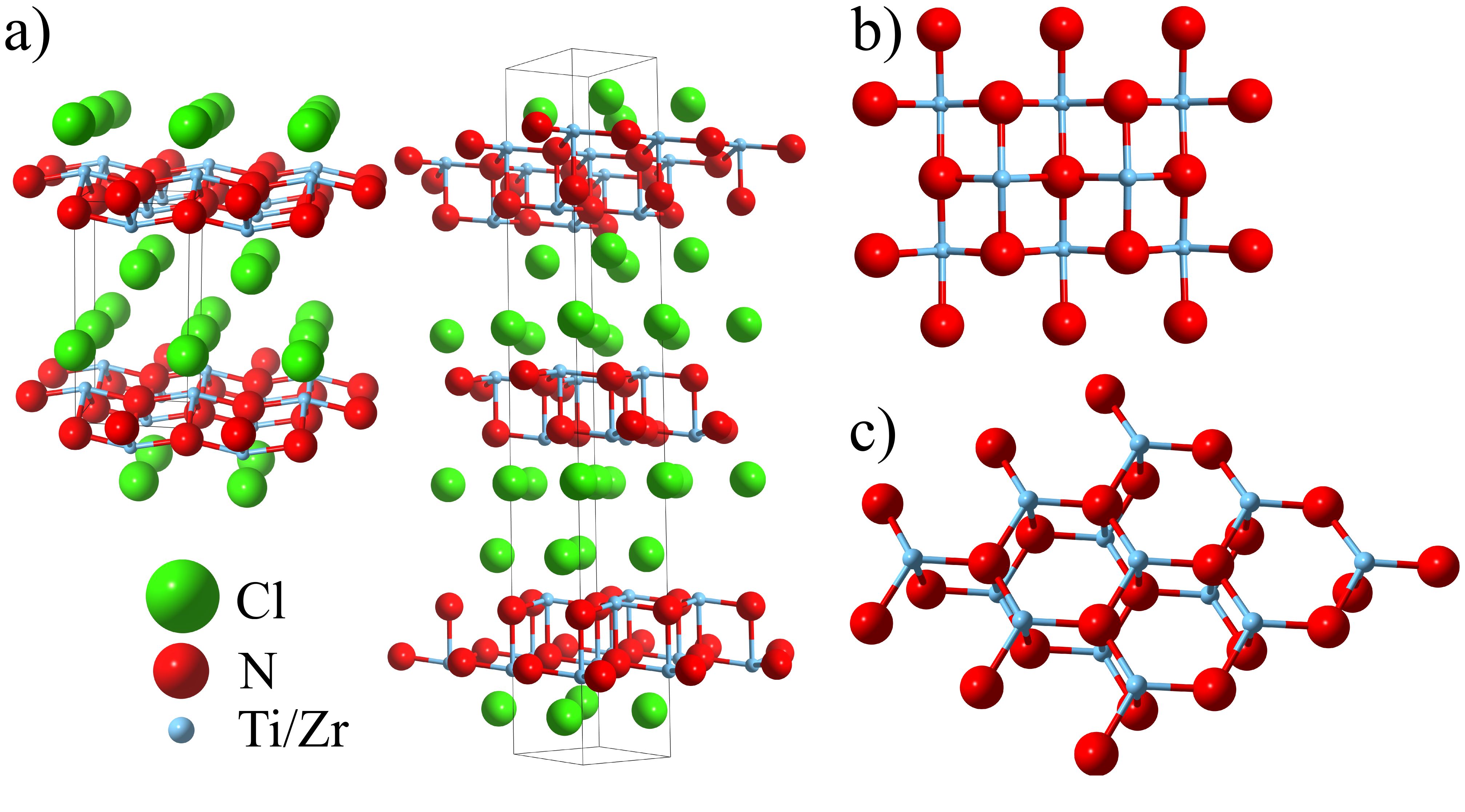}
  \caption{a) Structure of $\alpha$-TiNCl (left) and structure of  $\beta$-ZrNCl (right). b) View
from top on the Ti-N lattice. c) View from top on the Zr-N lattice}
\label{mncl_fig1}

\end{figure}

%%%%%%%%%%%%%%%%%%%%%%%%%%%%%%%%%%%%%%%%%%%%%%%%%%%%%%%%%%%%%%%%%%%%%%
$\alpha$-TiNCl crystallizes in the orthorhombic space group P${mmn}$ with the
FeOCl structure type ($\alpha$ structure), which is related to the PbFCl structure by buckling the
anion layer. The PbFCl structure type is known from the Fe based high T$_c$ compounds such as LiFeAs, which adopts this structure type. In contrast, $\beta$-ZrNCl, crystallizes in the
rhombohedral space group R${\bar{3}m}$ within the SmSI structure type (Figure
\ref{mncl_fig1} a) which can be viewed as a stuffing of the ZrCl structure with
N-atoms, resulting in differently stacked Cl-Zr$_2$N$_2$-Cl layers ($\beta$ structure). The ZrCl structure basically consists of different stacking of NiAs-type layers.
\cite{juza196,schurz2011} \\
In both structure types, each \textit{M}-atom is surrounded by 4 N-atoms and coordinated by different amounts of Cl-atoms
above and below the layer. The lattice spanned by the \textit{M}N sheets in
TiNCl is rectangular whereas the lattice in ZrNCl is a double honeycomb layer
(Figure \ref{mncl_fig1} b and c). The local geometry around the \textit{M}-atoms
is thus very similar in contrast to the global geometry of the lattice.
In ZrNCl, intercalation of Li leads to a change in the crystal structure
by changing the stacking sequence of Cl-Zr$_2$N$_2$-Cl layers, however the space
group remains the same. During this change of stacking, the Zr-Zr distance within a layer
decreases from d(Zr-Zr) = 3.34 \AA{} to d(Zr-Zr) = 3.08 \AA{} in
Li$_{0.16}$ZrNCl, which is lower than the Zr-Zr distance of 3.18 \AA{} in
elemental Zr and is well comparable  to the Zr-Zr distance in ZrCl with 3.08
\AA{}. \cite{shamoto1998,adolphson1976}
% Comment by Lukas: Zr is superconducting below 2K, however under pressure T$_c$
%goes up to 10-15 K!
Li-intercalation in $\alpha$-TiNCl leads to no change in the layer stacking,
however the Ti-Ti distance d(Ti-Ti) decreases from 3.00 \AA{} to 2.88 \AA{} in
Li$_{0.16}$TiNCl (2.86 \AA{} in elemental Ti), showing similar behavior as ZrNCl.
\cite{yamanaka2009}
%%%%%%%%%%%%%%%%%%%%%%%%%%%%%%%%%%%%%%%%%%%%%%%%%%%%%%%%%%%%%%%%%%%%%%
\begin{figure}[!htbp,]
\centering
  \includegraphics[width=0.5\textwidth]{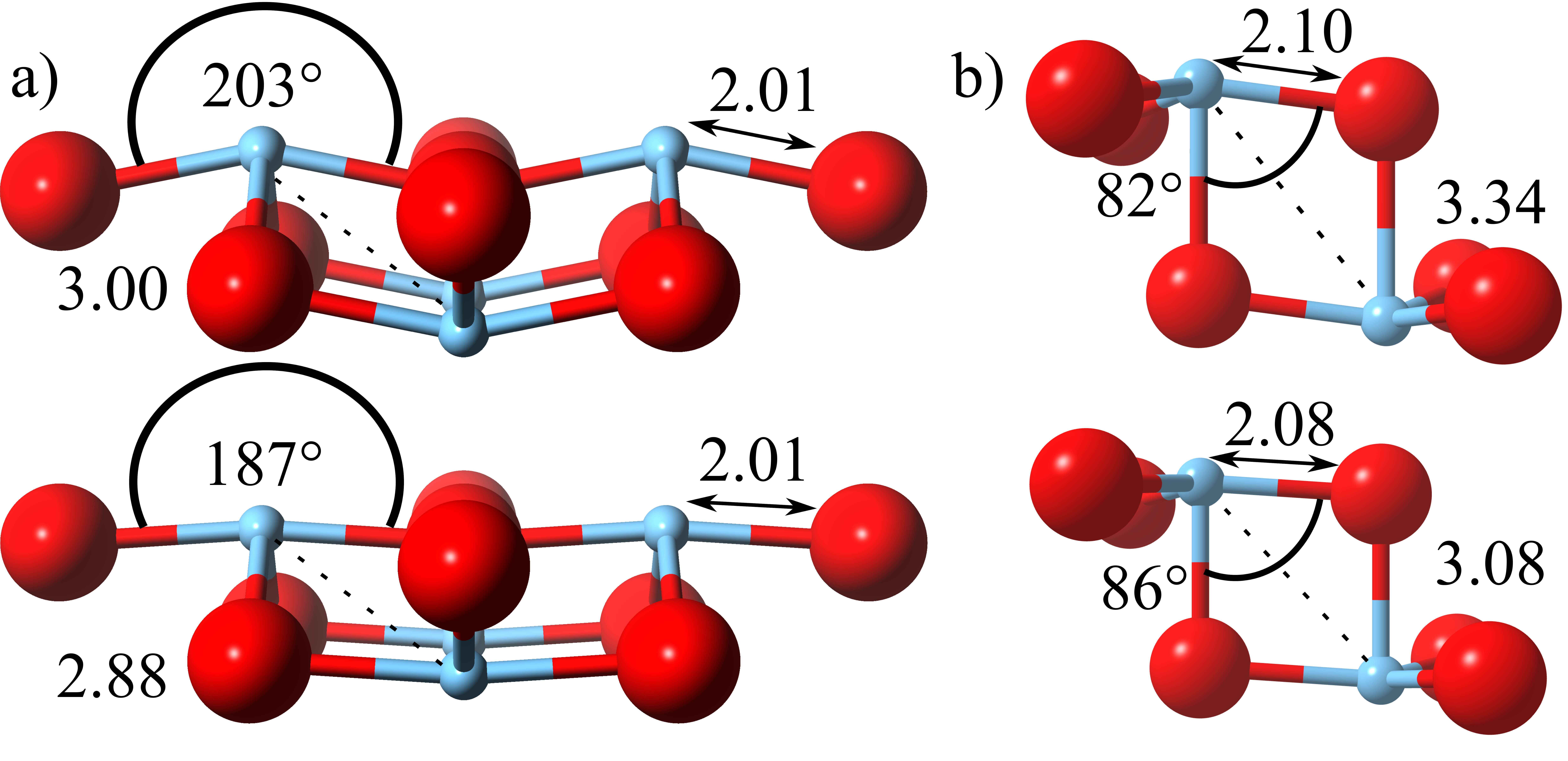}
  \caption{Structural changes upon doping for TiNCl (a) and ZrNCl (b). Data taken from Ref. \cite{schurz2011}}
\label{mncl_fig2}
\end{figure}
%%%%%%%%%%%%%%%%%%%%%%%%%%%%%%%%%%%%%%%%%%%%%%%%%%%%%%%%%%%%%%%%%%%%%%

\section{Methods of calculation}

The calculations were performed in the framework of DFT using the \textsc{wien2k} \cite{blaha2001} code with a full-potential
linearized augmented plane-wave and local orbitals [FP-LAPW + lo] basis
\cite{singh2006,madsen2001,sjaestedt_alternative_2000} together with the Perdew
Burke Ernzerhof (PBE) parametrization \cite{perdew_generalized_1996} of the
generalized gradient approximation (GGA) as the
exchange-correlation functional. The plane wave cut-off parameter
R$_{MT}$K$_{MAX}$ was set to 7 and the irreducible Brillouin zone was sampled by
100-120 k-points in the insulating compounds and by 498 to 600 in the doped
metallic compounds depending on the $\alpha$ or $\beta$ crystal structure.
Experimental lattice constants were used and the atomic positions were optimized
by minimization of the forces, which agree well with the reported ones.
Li doping has been simulated by the virtual crystal approximation (VCA) by positioning
Li on the positions suggested by neutron scattering and removing charge on the
Li atoms. We choose to simulate a Li doping corresponding to compounds with the formula Li$_{0.25}$\textit{M}NCl, since T$_c$ is roughly independent of the amount of Li intercalated.

\section{Results and Discussion}
%%%%%%%%%%%%%%%%%%%%%%%%%%%%%%%%%%%%%%%%%%%%%%%%%%%%%%%%%%%%%%%%%%%%%%
\begin{figure}[!htbp,]
\centering
  \includegraphics[width=0.85\textwidth]{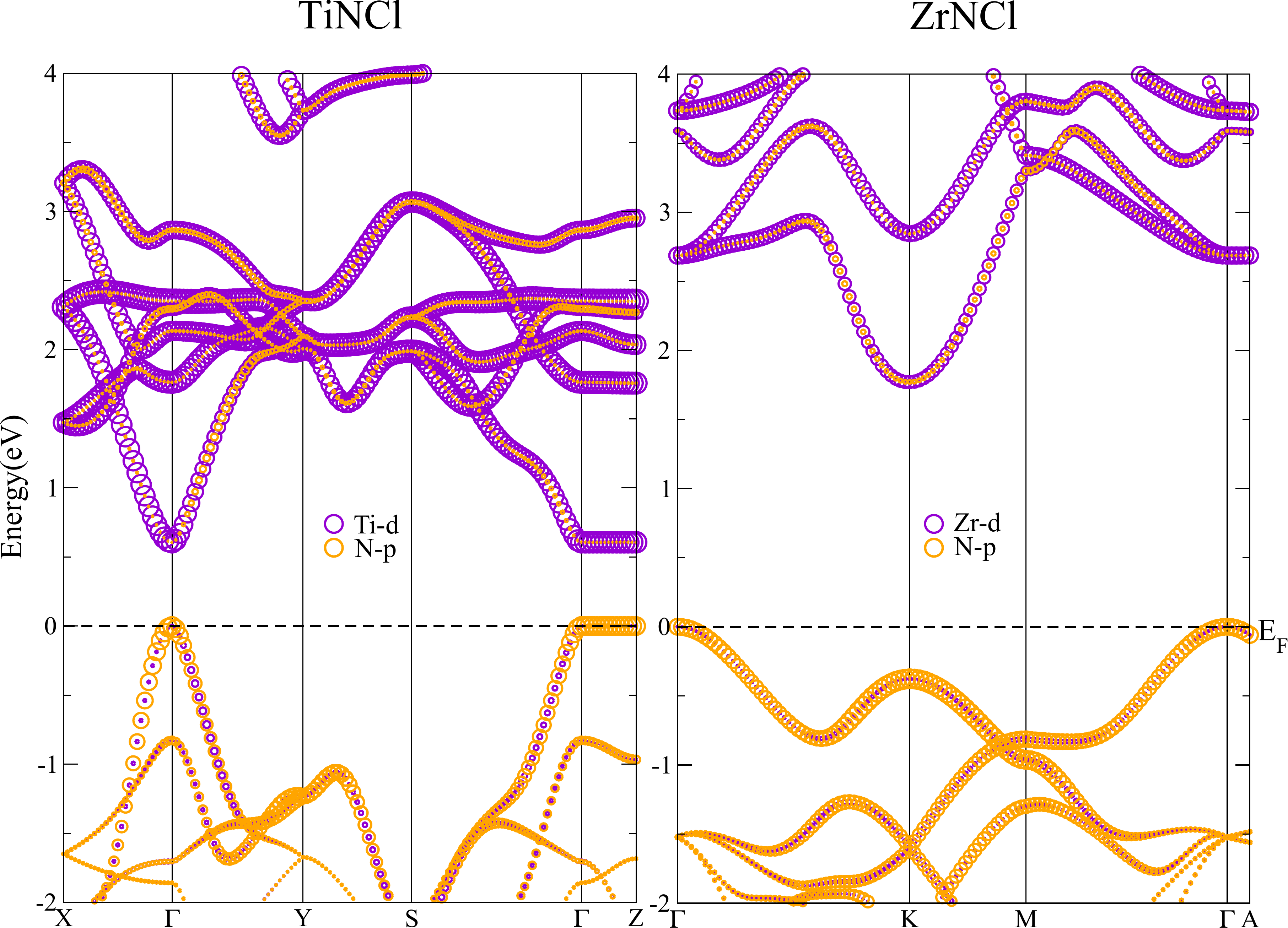}
  \caption{Band structures of $\alpha$-TiNCl and $\beta$-ZrNCl. The
size of the dots indicate the contribution from the atomic-orbitals.
Ti/Zr-\textit{d} states are purple, N-\textit{p} states orange.
}
\label{mncl_fig3}
\end{figure}
%%%%%%%%%%%%%%%%%%%%%%%%%%%%%%%%%%%%%%%%%%%%%%%%%%%%%%%%%%%%%%%%%%%%%%
%%%%%%%%%%%%%%%%%%%%%%%%%%%%%%%%%%%%%%%%%%%%%%%%%%%%%%%%%%%%%%%%%%%%%%
\begin{figure}[!htbp,]
\centering
  \includegraphics[width=0.85\textwidth]{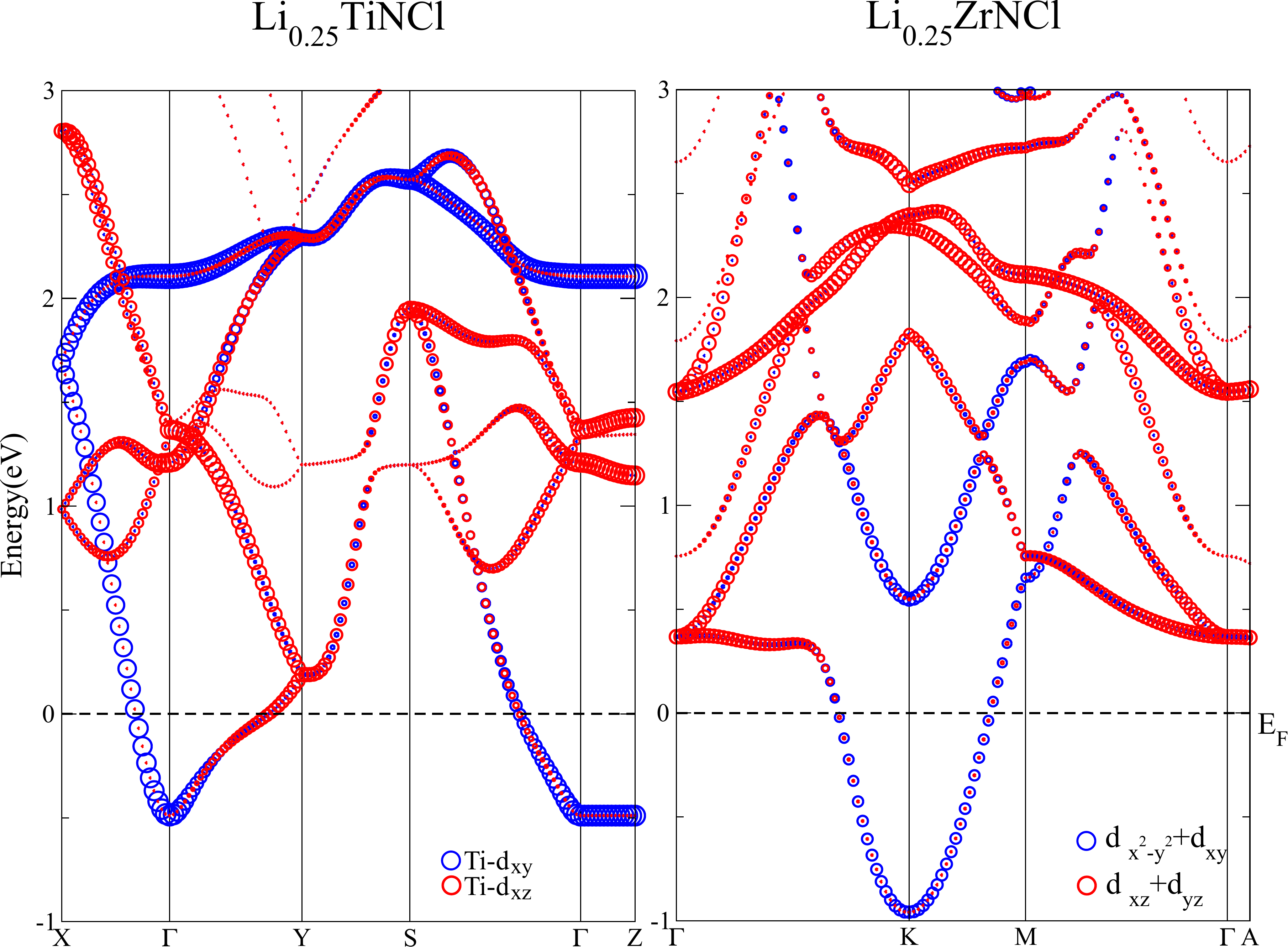}
  \caption{Band structures of $\alpha$-Li$_{0.25}$TiNCl and $\beta$-Li$_{0.25}$ZrNCl. The
size of the dots indicate the contribution from the atomic-orbitals.
Ti-\textit{d}$_{xy}$ are blue and Ti-\textit{d}$_{xz}$ + \textit{d}$_{yz}$ are
red in the left panel. Zr-\textit{d}$_{x^2-y^2}$+ \textit{d}$_{xy}$ are blue and Zr-\textit{d}$_{xz}$ +
\textit{d}$_{yz}$ are red in the right panel}
\label{mncl_fig4}
\end{figure}
%%%%%%%%%%%%%%%%%%%%%%%%%%%%%%%%%%%%%%%%%%%%%%%%%%%%%%%%%%%%%%%%%%%%%%
In TiNCl, [TiN]$^+$ layers dominate the electronic structure around the Fermi level
(Figure \ref{mncl_fig3}). The compound is semiconducting with a direct band gap of 0.6 eV at the $\Gamma$-point. The conduction band consists
mostly of N-\textit{p} states and the valence band is a single
Ti-\textit{d}$_{xy}$ band slightly hybridized with N-\textit{p} states.
Upon intercalation of Li, the band structure behaves in a non-rigid band 
manner, shifting Ti-\textit{d}$_{xz}$ states closer to the Fermi level at the
Y-point (Figure \ref{mncl_fig4}). The decrease in Ti-Ti distance leads to an increase of bandwidth of the
Metal-\textit{d} bands upon doping, consistent with an increase of metal-metal
interaction.
ZrNCl is a band insulator with an indirect gap of 2 eV from the $\Gamma$- to the K-point (Figure \ref{mncl_fig3}).
The valence band consists mostly of N-\textit{p} states hybridized with
Zr-\textit{d} states and the conduction band consists of Zr-\textit{d}$_{x^2-y^2}$
+ \textit{d}$_{xy}$ bands hybridized with N-\textit{p} states.
In accordance with the crystal structure one can think of covalent
closed shell [ZrN]$^+$ layers. Electron doping with Li shifts the Fermi level
into a single conduction band with strong Zr-\textit{d}-N-\textit{p} hybridization (Figure \ref{mncl_fig4}). Li
intercalation thus leads to [ZrN]$^{(1-x)+}$ layers with Zr in a \textit{d}$^{x}$
configuration where \textit{x} is the amount of Li intercalated. This is
accompanied by a large decrease in Zr-Zr distance, indicating strong Zr-Zr
interactions. \\
This is consistent with previous calculations that indicate that the single conduction band
stems from bonding metal-\textit{d} interactions. \cite{claudia,istomin1999}
The \textit{d}$^{x}$
configuration on the \textit{M}-atom is in proximity to a \textit{d}$^{1}$
configuration that has been shown to be favorable for superconductivity in other layered transition metal compounds. \cite{gabovich2001charge} \\
%%%%%%%%%%%%%%%%%%%%%%%%%%%%%%%%%%%%%%%%%%%%%%%%%%%%%%%%%%%%%%%%%%%%%%
\begin{figure}[!htbp,]
\centering
  \includegraphics[width=0.5\textwidth]{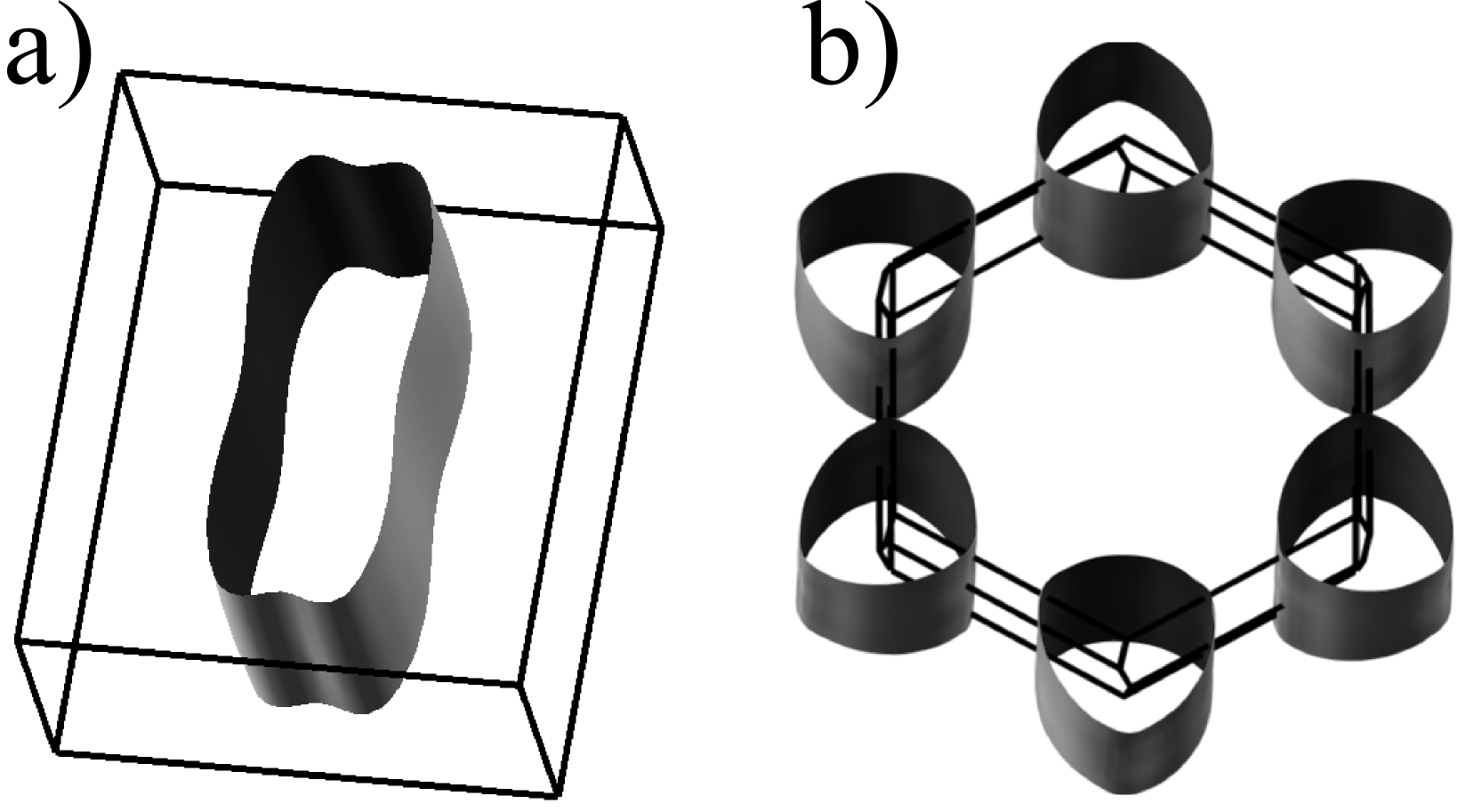}
  \caption{Fermi surfaces for the Li intercalated compounds. a)
Li$_{0.25}$TiNCl, b) Li$_{0.25}$ZrNCl}
\label{mncl_fig5}
\end{figure}
%%%%%%%%%%%%%%%%%%%%%%%%%%%%%%%%%%%%%%%%%%%%%%%%%%%%%%%%%%%%%%%%%%%%%%
The band structures confirm the similarities in bonding between the two
crystallographically different compounds. In both structures, the conduction
band consists of \textit{M}-\textit{d}-bands slightly hybridized with
N-\textit{p} states. Upon doping the \textit{M}-\textit{M} distance decreases by
a large amount, which together with the band character at the Fermi level
indicates strong \textit{M}-\textit{M} interactions .\\
The Fermi surfaces of the doped compounds have
similar features (Figure \ref{mncl_fig5}) and indicate a
2-dimensional electronic structure with little dispersion along the z-direction.

\begin{figure}[!htbp,]
  \centering
  \includegraphics[width=0.85\textwidth]{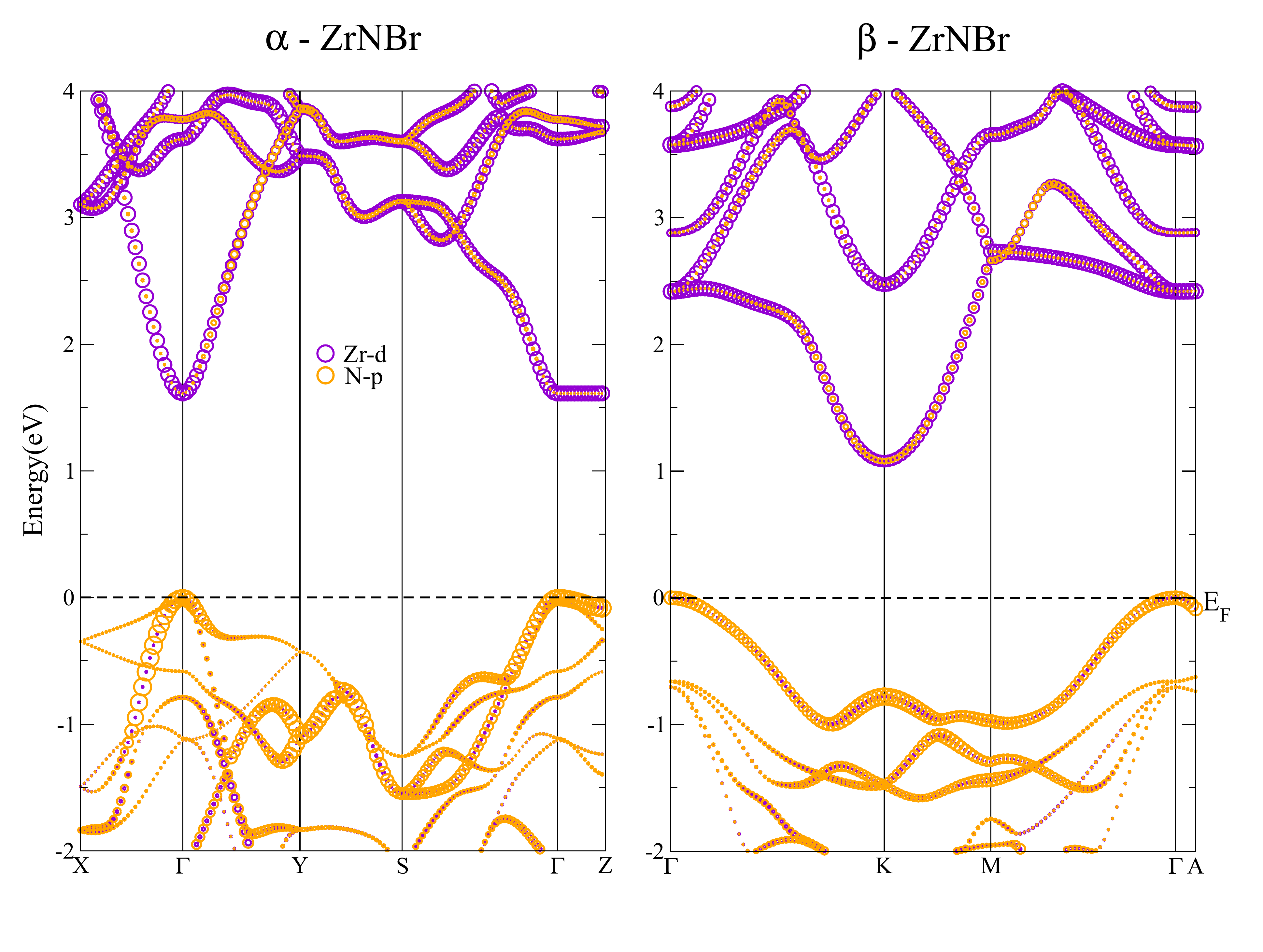}
 \caption{Band structures of $\alpha$ and $\beta$-ZrNBr. The
size of the dots indicate the contribution from the atomic-orbitals.
Zr-\textit{d} states are purple, N-\textit{p} states orange.}
\label{mncl_fig6}
\end{figure}

For TiNCl the Fermi surface consists of slightly warped rectangular sheets
whereas the Fermi surface of ZrNCl shows 6 trigonally distorted cylinders at the
zone boundaries.
Both Fermi surfaces reflect the underlying symmetry of the lattice and show
tendencies towards nesting due to the lack of dispersion along the z direction.
\\
Thus, the electronic structures of both compounds are very similar,
despite the different crystal symmetry. The band structures show the importance
of local bonding whereas both Fermi surfaces show the proximity towards
electronic instabilities due to nesting. \\
This idea is supported by the fact that ZrNBr, which can be synthesized in both the $\alpha$ and $\beta$-structure, becomes superconducting upon Li intercalation with comparable T$_c$'s. \cite{ZrNBr}
Figure \ref{mncl_fig6} shows the band structures of $\alpha$ and $\beta$-ZrNBr. Both band structures share the same features as the ones discussed above.
The conduction band mainly consists of Zr-\textit{d}  states, whereas the valence band stems mostly from N-\textit{p} states.
%%%%%%%%%%%%%%%%%%%%%%%%%%%%%%%%%%%%%%%%%%%%%%%%%%%%%%%%%%%%%%%%%%%%%
%%%%%%%%%%%%%%%%%%%%%%%%%%%%%%%%%%%%%%%%%%%%%%%%%%%%%%%%%%%%%%%%%%%%%%
The T$_c$ of this class of compounds is not very dependent on the amount of
doping and one might speculate that the change in \textit{M}-\textit{M} distance
is the most important factor for the appearance of superconductivity. \cite{yamanaka2010}
After a critical doping level necessary to induce the shift in \textit{M}
distance, T$_c$ is more or less independent of the dopant concentration. \cite{PhysRevLett.97.107001}   \\
The proximity to a \textit{d}$^1$ configuration of the metal allows for the
comparison with other layered early transition metal chalcogenides that become
superconducting upon doping as well, such as TaSe$_2$ or Spinels as LiTi$_2$O$_4$.
\cite{gabovich2001charge,johnston1973LiTiO} The crystal structures of these
compounds also allow for direct \textit{M}-\textit{M} interaction due to corner
sharing octahedra or trigonal prisms. 
TiNCl can also be described as sheets of distorted edge sharing
TiN$_4$Cl$_2$ octahedra. Upon Li-intercalation the
distortion reduces, due to the reduction in Ti-Ti distance. We conjecture that
it is the local bonding and symmetry that is important for superconductivity
for this class of materials together with structure types that have the
necessary flexibility. The observation of close \textit{M}-\textit{M} distances
has also been made in \textit{Ln}Ni$_2$B$_2$C, during the discussion of the importance of
phonons for superconductivity in this class of compounds.
\cite{Loureiro2001675,mattheiss1994lf,simon1996} Furthermore, the iron based
high-T$_c$ superconductors crystallize in derivatives of the anti-PbFCl
structure, with the structural motive of edge sharing tetrahedra, which allows
for strong \textit{M}-\textit{M} interactions and leads to a complex magnetic
phenomenology. \cite{johrendt2011structural} As an example, the T$_c$ of the
$RE$OFeAs systems is highest 
when the angle of the edge sharing tetrahedra is closest to the perfect
tetrahedral angle. \cite{lee2008effect,mizuguchi2010anion}  Another example is
the superconducting non-magnetic structural analog of LiFeAs, NaAlSi, where a
short Al-Al distance leads to a very dispersive band close to the Fermi level.
\cite{schoop2012effect}

\section{Conclusion}
We showed that there is a close connection between the electronic structure of
$\alpha$-TiNCl and $\beta$-ZrNCl, despite their different lattice symmetries,
i.e. orthorhombic and hexagonal. Both compounds show strong in plane
\textit{M}-N bonding and are band insulators. Upon doping the
\textit{M}-\textit{M} distance decreases, which leads to \textit{M}-\textit{M}
interactions that give rise to a single band composed of mostly in plane metal
\textit{d}-states that shows a tendency towards nesting, which is not surprising
due to the similarities of the compounds to known charge density wave systems. 
\cite{gabovich2001charge} We conjecture that the local bonding properties in
connection with certain structural motives and the vicinity of an electronic
instability, allow for the most propably unconventional superconductivity in this class of compounds. The example of ZrNBr which is superconducting in both $\alpha$ and $\beta$ structures further supports this idea.
It is this proximity to a charge density wave instability that makes this class of compounds special and distinguishes them from the cuprates where the mechanism is most probably spin-mediated. 
This can be put in perspective with a larger class of known superconductors such as early
transition-metal chalcogenides, spinels, borocarbides and even Fe-based
superconductors that show similar fingerprints in 
their electronic structure and bonding motives.

%%%%%%%%%%%%%%%%%%%%%%%%%%%%%%%%%%%%%%%%%%%%%%%%%%%%%%%%%%%%%%%%%%%%%%
\begin{ack}
We would like to acknowledge helpful discussions with Robert J. Cava.
This work was supported by the DFG within the SPP 1458.
\end{ack}
\bigskip
\bibliographystyle{iopart-num}
\bibliography{muechler_MNCL}

\providecommand{\newblock}{}
\begin{thebibliography}{10}
\expandafter\ifx\csname url\endcsname\relax
  \def\url#1{{\tt #1}}\fi
\expandafter\ifx\csname urlprefix\endcsname\relax\def\urlprefix{URL }\fi
\providecommand{\eprint}[2][]{\url{#2}}
% Bibliography created with iopart-num v2.1
% /biblio/bibtex/contrib/iopart-num

\bibitem{yamanaka1998superconductivity}
Yamanaka S, Hotehama K~i and Kawaji H {1998} {\em Nature\/} {\bf 392} 580--582

\bibitem{yamanaka2010}
Yamanaka S 2010 {\em Journal of Materials Chemistry\/} {\bf 20} 2922--2933

\bibitem{schurz2011}
Schurz C~M, Shlyk L, Schleid T and Niewa R 2011 {\em Zeitschrift f{\"u}r
  Kristallographie\/} {\bf 226} 395--416

\bibitem{yamanaka2009}
Yamanaka S, Yasunaga T, Yamaguchi K and Tagawa M 2009 {\em {Journal of
  Materials Chemistry}\/} {\bf 19} 2573--2582

\bibitem{PhysRevB.67.100509}
Tou H, Maniwa Y and Yamanaka S 2003 {\em Phys. Rev. B\/} {\bf 67}(10) 100509

\bibitem{kitora2007probing}
Kitora A, Taguchi Y and Iwasa Y 2007 {\em Journal of the Physical Society of
  Japan\/} {\bf 76}

\bibitem{PhysRevLett.97.107001}
Taguchi Y, Kitora A and Iwasa Y 2006 {\em Phys. Rev. Lett.\/} {\bf 97}(10)
  107001

\bibitem{PhysRevB.86.054513}
Akashi R, Nakamura K, Arita R and Imada M 2012 {\em Phys. Rev. B\/} {\bf 86}(5)
  054513

\bibitem{Kotliar2013}
Yin Z~P, Kutepov A and Kotliar G 2013 {\em Phys. Rev. X\/} {\bf 3}(2) 021011

\bibitem{istomin1999}
Istomin S~Y, K{\"o}hler J and Simon A 1999 {\em Physica C: Superconductivity\/}
  {\bf 319} 219--228

\bibitem{woodward1998electronic}
Woodward P and Vogt T 1998 {\em Journal of Solid State Chemistry\/} {\bf 138}
  207--219

\bibitem{PhysRevB.83.014509}
Yin Q, Ylvisaker E~R and Pickett W~E 2011 {\em Phys. Rev. B\/} {\bf 83}(1)
  014509

\bibitem{juza196}
Juza R and Heners J 1964 {\em Zeitschrift f{\"u}r anorganische und allgemeine
  Chemie\/} {\bf 332} 159--172

\bibitem{shamoto1998}
Shamoto S, Kato T, Ono Y, Miyazaki Y, Ohoyama K, Ohashi M, Yamaguchi Y and
  Kajitani T 1998 {\em {Physica C: Superconductivity}\/} {\bf 306}

\bibitem{adolphson1976}
Adolphson D~G and Corbett J~D 1976 {\em {Inorganic chemistry}\/}

\bibitem{blaha2001}
Blaha P, Schwarz K, Madsen G, Kvasnicka D and Luitz J 2001 {\em WIEN2k, An
  Augmented Plane Wave+ Local Orbitals Program for calculating Crystal
  Properties, Technische Universit{\"a}t Wien, Austria\/}

\bibitem{singh2006}
{D J Singh} and Nordstr{\"o}m L 2006 {\em Planewaves, Pseudopotentials, and the
  LAPW Method, Springer, New York, 2nd ed.\/}

\bibitem{madsen2001}
Madsen G~K~H, Blaha P, Schwarz K, Sj{\"o}stedt E and Nordstr{\"o}m L 2001 {\em
  Physical Review B\/}  195134

\bibitem{sjaestedt_alternative_2000}
Sj{\"o}stedt E, Nordstr{\"o}m L and Singh D~J 2000 {\em {Solid State
  Communications}\/} {\bf 114} 15--20 ISSN 0038-1098

\bibitem{perdew_generalized_1996}
Perdew J~P, Burke K and Ernzerhof M 1996 {\em Physical Review Letters\/} {\bf
  77} 3865

\bibitem{claudia}
Felser C and Seshadri R 1999 {\em J. Mater. Chem.\/} {\bf 9}(2) 459--464

\bibitem{gabovich2001charge}
Gabovich A, Voitenko A, Annett J and Ausloos M 2001 {\em Superconductor Science
  and Technology\/} {\bf 14} R1

\bibitem{ZrNBr}
Fogg A~M, Green V~M and O'Hare D 1999 {\em Journal of Materials Chemistry\/}
  {\bf 9} 1547--1551

\bibitem{johnston1973LiTiO}
Johnston D, Prakash H, Zachariasen W and Viswanathan R 1973 {\em Materials
  Research Bulletin\/} {\bf 8} 777--784

\bibitem{Loureiro2001675}
Loureiro S, Kealhofer C, Felser C and Cava R 2001 {\em Solid State
  Communications\/} {\bf 119} 675 -- 679 ISSN 0038-1098

\bibitem{mattheiss1994lf}
Mattheiss L, Siegrist T and Cava R 1994 {\em Solid State Communications\/} {\bf
  91} 587--590

\bibitem{simon1996}
Simon A, B{\"a}cker M, Henn R, Felser C, Kremer R, Mattausch H and Yoshiasa A
  1996 {\em Zeitschrift f{\"u}r anorganische und allgemeine Chemie\/} {\bf 622}
  123--137

\bibitem{johrendt2011structural}
Johrendt D, Hosono H, Hoffmann R~D and P{\"o}ttgen R 2011 {\em Zeitschrift
  f{\"u}r Kristallographie\/} {\bf 226} 435--446

\bibitem{lee2008effect}
Lee C~H, Iyo A, Eisaki H, Kito H, Fernandez-Diaz M~T, Ito T, Kihou K, Matsuhata
  H, Braden M and Yamada K 2008 {\em Journal of the Physical Society of
  Japan\/} {\bf 77}

\bibitem{mizuguchi2010anion}
Mizuguchi Y, Hara Y, Deguchi K, Tsuda S, Yamaguchi T, Takeda K, Kotegawa H, Tou
  H and Takano Y 2010 {\em Superconductor Science and Technology\/} {\bf 23}
  054013

\bibitem{schoop2012effect}
Schoop L, M{\"u}chler L, Schmitt J, Ksenofontov V, Medvedev S, Nuss J, Casper
  F, Jansen M, Cava R and Felser C 2012 {\em Physical Review B\/} {\bf 86}
  174522

\end{thebibliography}

\end{document}